\documentclass[traditabstract]{aa}  

\usepackage{graphicx}                    
\usepackage{bm}
\usepackage{color}                       
\usepackage{graphicx}
\usepackage{xcolor}
\usepackage{bm}%
\usepackage{amsmath}
\usepackage{multirow}
\usepackage{subfigure}
\usepackage{booktabs}
\usepackage{graphicx}
\usepackage{amsfonts,amssymb}
\usepackage{times}

\usepackage{grffile}

\usepackage{hyperref}

\newcommand{\be}{\begin{equation}}
\newcommand{\ee}{\end{equation}}

\def\lsim{\lower.4ex\hbox{$\;\buildrel <\over{\scriptstyle\sim}\;$}}
\newcommand{\Ha}{\mbox{Ha}}
\newcommand{\Pm}{\mbox{Pm}}
\newcommand{\Rm}{\mbox{Rm}}
\newcommand{\Mm}{\mbox{Mm}}

\newcommand{\Sc}{\mbox{Sc}}
\newcommand{\Rey}{\mbox{Re}}

\def\Om{{\it \Omega}}
\def\R{{R\"udiger}}
\def\A{Alfv\'en}
\newcommand{\gsim}{\lower.7ex\hbox{$\;\stackrel{\textstyle>}{\sim}\;$}}

\renewcommand{\vec}[1]{\mbox{\boldmath$#1$}}
\def\gsim{\lower.4ex\hbox{$\;\buildrel >\over{\scriptstyle\sim}\;$}} 
\def\lsim{\lower.4ex\hbox{$\;\buildrel <\over{\scriptstyle\sim}\;$}}
\def\A{Alfv\'en}
\def \Om  {{\it \Omega}}
\begin{document}



\title{Mixing of a passive scalar by the instability of a differentially rotating axial pinch}
\titlerunning{Mixing of a passive scalar}
%
\author{A. Paredes \and M. Gellert \and G.~R\"udiger}
 \authorrunning{A. Paredes, M. Gellert, G.~R\"udiger}      


%
  \institute{  Leibniz-Institut f\"ur Astrophysik Potsdam, An der Sternwarte 16, D-14482 Potsdam, Germany,
                     \email{ aparedes@aip.de, mgellert@aip.de, gruediger@aip.de}
		     }

\date{Received; accepted}
 
\abstract{The mixing of a passive scalar like lithium,  beryllium or temperature 
fluctuations due to the magnetic Tayler instability of a 
rotating axial   pinch is considered.  Our study is carried out within a 
Taylor-Couette setup for two  rotation laws: quasi-Kepler and solid-body 
rotation.  The minimum  magnetic Prandtl number used is $0.05$  while the molecular 
Schmidt number $\Sc$ of the fluid varies between $0.1$ and $2$ where the latter 
value characterizes  the solar tachocline.  An effective diffusivity coefficient 
for the mixing  is numerically measured by the decay process of a global 
concentration peak located  between both cylinder walls.  We find that  only models 
with  $\Sc$ exceeding  0.1 do basically provide finite eddy diffusivity values.
We also find that for quasi-Kepler rotation  at a  magnetic Mach number 
$\Mm\simeq 2$ the flow transits from the slow-rotation regime to the  
fast-rotation regime  dominated by the Taylor-Proudman theorem.  Just for this latter
state, the eddy diffusivity  values peak in their dependence on the Reynolds number of the global rotation. 
For fixed Reynolds number the relation between the normalized eddy diffusivity 
and the Schmidt number of the fluid is always  linear so that also a linear 
relation between the eddy diffusivity and the molecular viscosity results  
exactly in the sense proposed by Schatzman  (1977). The numerical value of the coefficient in this 
relation, here called the Schatzman number, will reach a maximum at $\Mm\simeq 2$ and will rapidly decrease for large $\Mm$.
Because of the high values of the solar Hartmann number, the Schatzman number would strongly exceed  the value of about 100 needed  to describe the 
mixing below the solar convection zone.   After our  results, however,  the diffusion coefficient will be strongly suppressed  by the solar rotation  if its 
magnetic Mach number fulfills $\Mm\gg 1$ implying  that only toroidal  magnetic fields of order kG can exist in the solar tachocline.}


%
\keywords{Instabilities - Magnetic fields - Diffusion - Turbulence - Magnetohydrodynamics (MHD)}

\maketitle
\section{Introduction}
Some  of the open problems in the modern stellar physics are related to empirical findings    of  rotational periods.
Numerical simulations of supernova explosions  yield rotation periods of 1 ms 
for the newborn neutron stars but the observations provide periods which are 
much longer. Less compact objects like the White Dwarfs are predicted to have equatorial velocities higher than 
15 km/s, but observations show velocities lower than 10 km/s (Berger et al. 2005). Obviously, an outward transport of angular momentum 
happened which cannot be explained with the molecular values of the 
viscosity. As the cores of the progenitor of newborn neutron stars and White 
dwarfs are always stably stratified an alternative instability must exist in the 
far-developed main-sequence stars which transports angular momentum but which 
does not lead to a too intensive mixing, which seems incompatible with observations of surface abundances in massive stars (Brott et al. 2008).

Even solar observations lead to similar  conclusions. The present-day solar core 
rotates rigidly but the  convection zone still contains lithium  which after a 
diffusion process between  the convection zone and the burning region 40.000 km 
under its bottom is destroyed. In order to explain the lithium decay time of 
about 1....10 Gyr the effective diffusion coefficient must exceed the molecular 
viscosity  by one or two orders of magnitude. Schatzman (1969, 1977) suggested 
some sort of instability -- appearing in the stably stratified stellar radiation
zones -- as the generator of such   mild extra transport of passive chemicals. 
Lebreton \& Maeder (1987) considered for the mixing in the  solar model a 
relation $D^*={Re}^* \nu$ for the diffusion coefficient (after the notation of 
Schatzman, see Zahn 1990)  with ${ Re}^*\simeq 100$. This is a rather small 
value which leads  to   $D^*\lsim 10^3$ cm$^2$/s for the solar plasma shortly 
below the solar tachocline.  Brun et al. (1998) work with ${ Re}^*\simeq 20$ 
while a more refined model of the solar radiative zone starts from the molecular 
value  $D\simeq 10$ cm$^2$/s and reaches  $D^*\simeq 10^{2....4}$ under the 
influence of a heuristically  postulated hydrodynamical instability (Brun et al. 1999).

On the other hand, in order to reproduce  the rigid rotation of the solar 
interior the viscosity must exceed its molecular value by more than two orders 
of magnitude.  R\"udiger \& Kitchatinov (1996)  assume  that internal differential rotation 
interacts with a fossil poloidal magnetic field within the core.  The rotation 
becomes uniform along the field lines.   A problem is formed by the   `islands' 
of fast rotation  due to the non uniformity of the field lines. A  viscosity of 
more than  $10^{4} $ cm$^2$/s is needed to smooth out such artificial  peaks in 
the resulting rotation laws.

The   value of $3\cdot 10^4$~cm$^2$/s is  also reported by Eggenberger et al. (2012) to produce an internal rotation law of the red giant {KIC 8366239}  
consistent with asteroseismic results obtained by the {\sc Kepler}  
mission Beck et al. (2012). Also, Deheuvels et al. (2012, 2014)
derive  an internal rotation profile of  the early red giant KIC 7341231  with a 
core spinning at least five times faster than the surface. This is  less than  the 
stellar evolution codes yield  without extra angular momentum transport from the 
core  to the envelope Ceillier et al. (2012). 
  
That the effective viscosity is stronger  amplified than the effective 
diffusivity suggests a magnetic background of the phenomena. Magnetic 
fluctuations transport angular momentum via the Maxwell stress but they do not 
transport chemicals. Most sorts of MHD turbulence should thus provide higher 
eddy viscosity values rather than eddy diffusivity values, i.e. both Prandtl 
numbers as well as the Schmidt number must not exceed unity.

In the present paper the Tayler instability (Tayler 1957, 1973) under the presence of 
(differential) rotation is probed to produce diffusion coefficients for passive 
scalars. By linear theory the instability map is obtained for the unstable 
nonaxisymmetric mode with $m=1$. The eigenvalue problem is formulated for a 
cylindrical Taylor-Couette container where the gap between both rotating 
cylinders is filled with a conducting fluid of given magnetic Prandtl number. 
Inside the cylinders homogeneous axial electric-currents exist which produce an 
azimuthal magnetic field with the fixed radial profile $B_\phi\propto R$ which 
-- if strong enough -- is unstable even without rotation. It is known  that for 
magnetic Prandtl number of order unity a rigid rotation strongly suppresses the 
instability but -- as we shall show -- a differential rotation with negative 
shear re-destabilizes the flow so that a wide domain exists in the instability 
map wherein the  nonlinear code provides the spectra of the  flow and field 
fluctuations.  Between the rotating cylinders a steep radial profile  for the 
concentration of a passive scalar is initially established which decays in time by the 
action of the flow fluctuations. The decay time is then  determined in order to 
find the diffusion coefficient.

\section{The rotating pinch}
In a Taylor-Couette setup, a fluid with   microscopic viscosity $\nu$ and  
magnetic diffusivity $\eta=1/\mu_0\sigma$ ($\sigma$ the electric conductivity) 
and a homogeneous axial 
current $\vec{J}=\textrm{curl}\, \vec{B}$ are considered. The  equations of the system are
\be
\frac{\partial \vec{U}}{\partial t} + (\vec{U} \cdot \nabla )\vec{U} = \frac{1}{\rho} \nabla P + \nu  \Delta \vec{U} 
+ \frac{1}{\mu_{0}\rho} \textrm{curl}\, \vec{B} \times \vec{B} \label{eq1},
\ee
\be
\frac{\partial \vec{B}}{\partial t} = \textrm{curl}\, (\vec{U} \times \vec{B}) + \eta \Delta \vec{B}  \label{eq2}
\ee
with $\textrm{div}\,\vec{U} =  \textrm{div}\,\vec{B} =0$  where $\vec{U}$ is the 
actual velocity, $\vec{B}$ the magnetic field and $P$ the pressure.  Their 
actual values may be split by $\vec{U}=\vec{\bar U}+ \vec{u}$, and 
accordingly for $\vec B$ and the pressure. The general solution of the 
stationary 
and axisymmetric equations is 
\be
{\bar U}_\phi=R\Om=a R+\frac{b}{R} , \quad {\bar U}_{r}={\bar U}_{z}=0,
\ee
\be
{\bar B}_\phi=A R, \quad {\bar B}_{r}={\bar B}_{z}=0  \label{basic}
\ee
with 
\be
a=\frac{\mu-r_{\rm in}^{2}}{1-r_{\rm in}^{2}}\Om_{\rm in}, \quad b=\frac{1-\mu}{1-r_{\rm in}^{2}}\Om_{\rm in}R_{\rm in}^{2} 
\ee
and with $
A={B_{\rm in}}/{R_{\rm in}}$.  Here   $a$ and  $b$  are constants and  $A$ represents the applied electric-current.  The rotating pinch is formed by  a uniform and  axial  mean-field 
electric-current. 
The solutions ${\bar U}_\phi$ and ${\bar B}_\phi$  are governed by the ratios
\begin{equation}
r_{\rm in}=\frac{R_{\rm{in}}}{R_{\rm{out}}}, \quad \mu=\frac{\Om_{\rm{out}}}{\Om_{\rm{in}}}, 
\label{mu}
\end{equation}
where $R_{\rm{in}}$ and $R_{\rm{out}}$ are the radii of the inner and the outer cylinder, $\Om_{\rm{in}}$ and $\Om_{\rm{out}}$ are their rotation rates.

Equations (\ref{eq1})--(\ref{eq2}) in its dimensionless  form become 
\begin{equation}
\Rey \Big(\frac{\partial \vec{U}}{\partial t} + (\vec{U}\cdot \nabla)\vec{U}\Big) = - \nabla P +  \Delta \vec{U} +  \Ha^2{\textrm{curl}}\, \vec{B} \times \vec{B} \nonumber,
\end{equation}
\begin{equation}
\Rm\Big( \frac{\partial \vec{B}}{\partial t}- {\textrm{curl}} (\vec{U} \times \vec{B})\Big)=  \Delta\vec{B} \label{mhd} 
\end{equation}
and $ {\textrm{div}}\, \vec{U} = {\textrm{div}}\, \vec{B} = 0$.
These equations are  numerically solved for no-slip boundary conditions and for  perfect-conducting cylinders which are 
unbounded in axial direction.  Those boundary conditions are applied at both $R_{\rm{in}}$ and $R_{\rm{out}}$.  The dimensionless free  parameters
in Eqs.~(\ref{mhd})  are  the Hartmann number ($\Ha$) and the Reynolds number ($\Rey$), 
\be
{\rm{Ha}}=\frac{B_{\rm{in}} R_0}{\sqrt{\mu_0 \rho \nu \eta}},  \quad \quad 
{\rm{Re}}=\frac{\Om_{\rm{in}} R_0^2}{\nu},
\label{def}
\ee
where $R_0=\sqrt{R_{\rm in}(R_{\rm{out}}-R_{\rm{in}})}$ is the unit of length
and $B_{\rm{in}}$   the azimuthal magnetic field at the inner cylinder.  With the magnetic Prandtl number 
\be
{\rm{Pm}} =\frac{\nu}{\eta}
\ee  
the magnetic Reynolds number of the 
rotation is $\Rm=\Pm\ \Rey$.  For the magnetic Prandtl number of the solar 
tachocline Gough (2007) gives the rather large value $\Pm\simeq 0.05$. There are 
even smaller numbers down to $\Pm=10^{-4}$ under discussion (see Brandenburg \& Subramanian 2005).  However, for the  aforementioned red giants one  
finds  $\Pm$ of the order of unity  (R\"udiger et al. 2014).
The code which solves the  equation system (\ref{mhd}) is   described  in 
Fournier et al. (2005)  where also the  detailed  formulation of the possible  boundary 
conditions can  be  found. For the present study only perfect-conducting 
boundaries have been considered.
\begin{figure}[h!]
\centering
\includegraphics[scale=0.40]{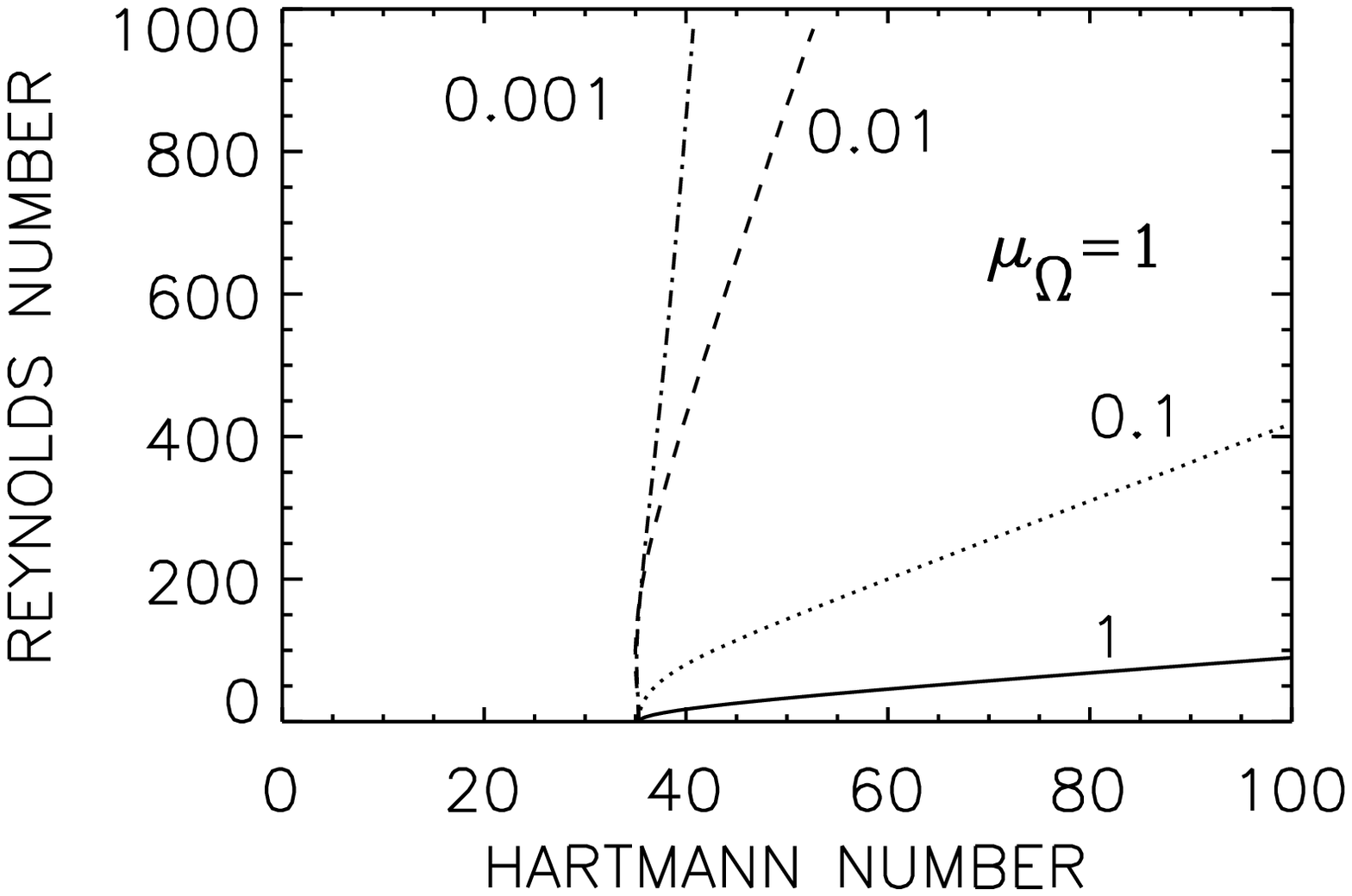}
\includegraphics[scale=0.40]{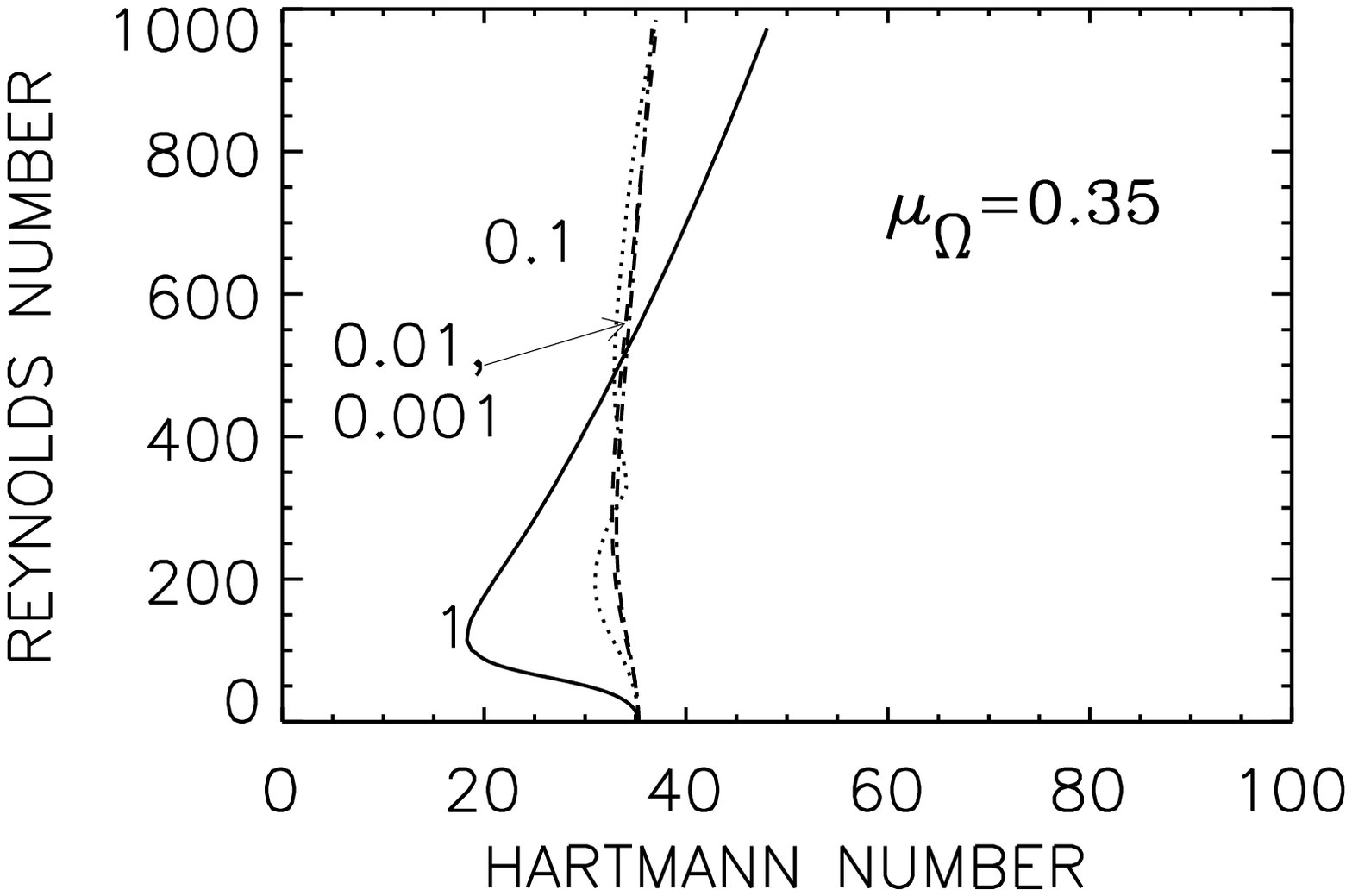}
\caption{Stability map for $m=1$ modes for  the pinch with  rigid rotation   
(top) and  with quasi-Kepler rotation law ($\mu=0.35$, bottom). The critical 
Hartmann number for resting cylinders is $\rm Ha=35.3$ for all $\Pm$. The curves 
are marked with their values of $\Pm$, the curve for $\Pm=0.001$ also represents 
all curves for smaller  $\Pm$.  $r_{\rm in}=0.5$, perfect-conducting 
boundaries.}
\label{f1}
\end{figure}

Figure \ref{f1} (top) shows the  map of marginal instability  for  the  rigidly 
rotating  pinch with $r_{\rm in}=0.5$  and for various $\Pm$.  The rotating 
fluid is unstable under the presence of a magnetic field with parameters on the right-hand 
side of the lines.  It also provides the influence of
 the magnetic Prandtl number on the rotational suppression. The $\Pm$-influence completely  disappears for 
the resting pinch  with $\Rey=0$. Note that the  rotating 
pinch  is massively stabilized for  magnetic Prandtl numbers  $\Pm\geq 0.1$.   
For very small magnetic Prandtl number the curves 
become indistinguishable, i.e. the marginal instability values under the 
influence of rigid rotation scale with $\Rey$ and $\Ha$ for $\Pm\to 0$.  This is 
a standard result for all linear  MHD equations in the induction-less 
approximation for $\Pm= 0$  (if such solutions exist). On the other hand, the 
rigidly rotating  pinch belongs to the configurations with the same radial 
profiles for velocity (here $\bar {U}_\phi\propto R$) and  magnetic field (here 
$\bar {B}_\phi\propto R$) defined by Chandrasekhar (1956). One can even show 
that all solutions  fulfilling this condition scale with $\Rey$ and $\Ha$ for 
$\Pm\to 0$ (R\"udiger et al. 2015a). These facts  imply that for a fixed magnetic resistivity    smaller molecular 
viscosities destabilize the rotating pinch.

The situation    changes   if the two cylinders are no longer rotating with the 
same angular velocity because also the shear energy is now able to excite 
nonaxisymmetric magnetic instability  patterns by interaction with 
toroidal fields  which are current-free  within the fluid. 
In this paper we shall present the results for the interaction
of shear with the azimuthal magnetic field which is due to the axial electric-current which defines the pinch.

The bottom panel of Fig.~\ref{f1} gives  the instability map for the magnetic 
instability of a fluid in quasi-Kepler rotation ($\mu=0.35$) for various  
$\Pm$.
It shows the influence of the differential rotation on the instability map of the rotating pinch. Again, of course,  the critical Hartmann
number for resting cylinders does not depend on the magnetic Prandtl number but in addition, the borderlines of the unstable region  for all $\Pm\leq 1$ do hardly differ.
For  the given  Reynolds number ranges,   the rotational suppression almost  disappears for $\Pm<1$.  For $\Pm=1$ and for $\Rey<400$ 
the instability becomes even subcritical and the rotational stabilization changes to a rotational destabilization.  
For too fast rotation, however, the subcritical excitation disappears but the rotational suppression is weaker than
 it is for rigid rotation. According to Fig.~\ref{f1} the value $\Pm=0.1$ which is mainly used in the 
calculations below already belongs to  the  small-$\Pm$ system.

\begin{figure}[h!]
\centering
\subfigure[Re = 500] {\includegraphics[height=0.22\textwidth,angle=-90]{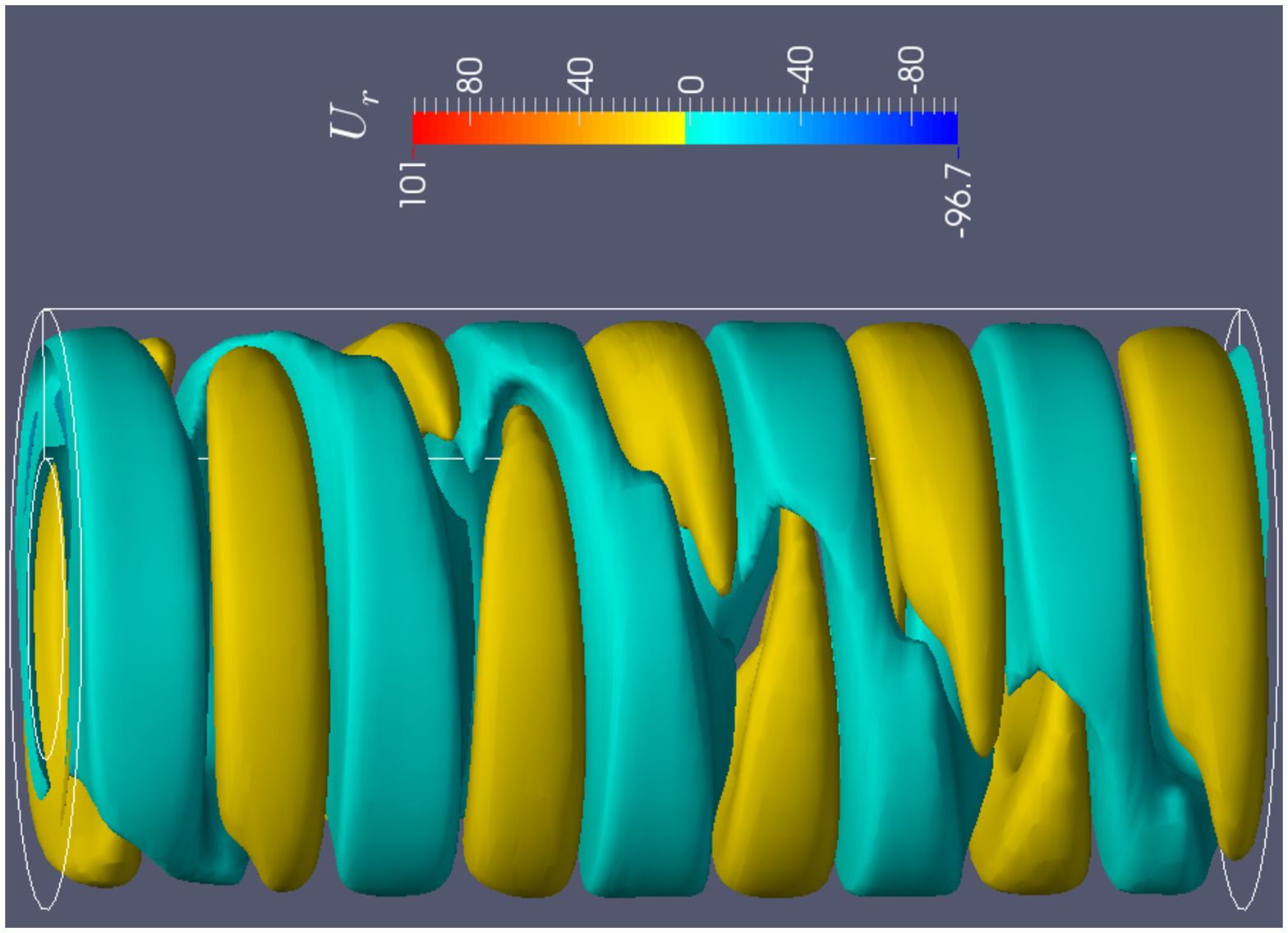}}\quad%
\subfigure[Re = 600] {\includegraphics[height=0.22\textwidth,angle=-90]{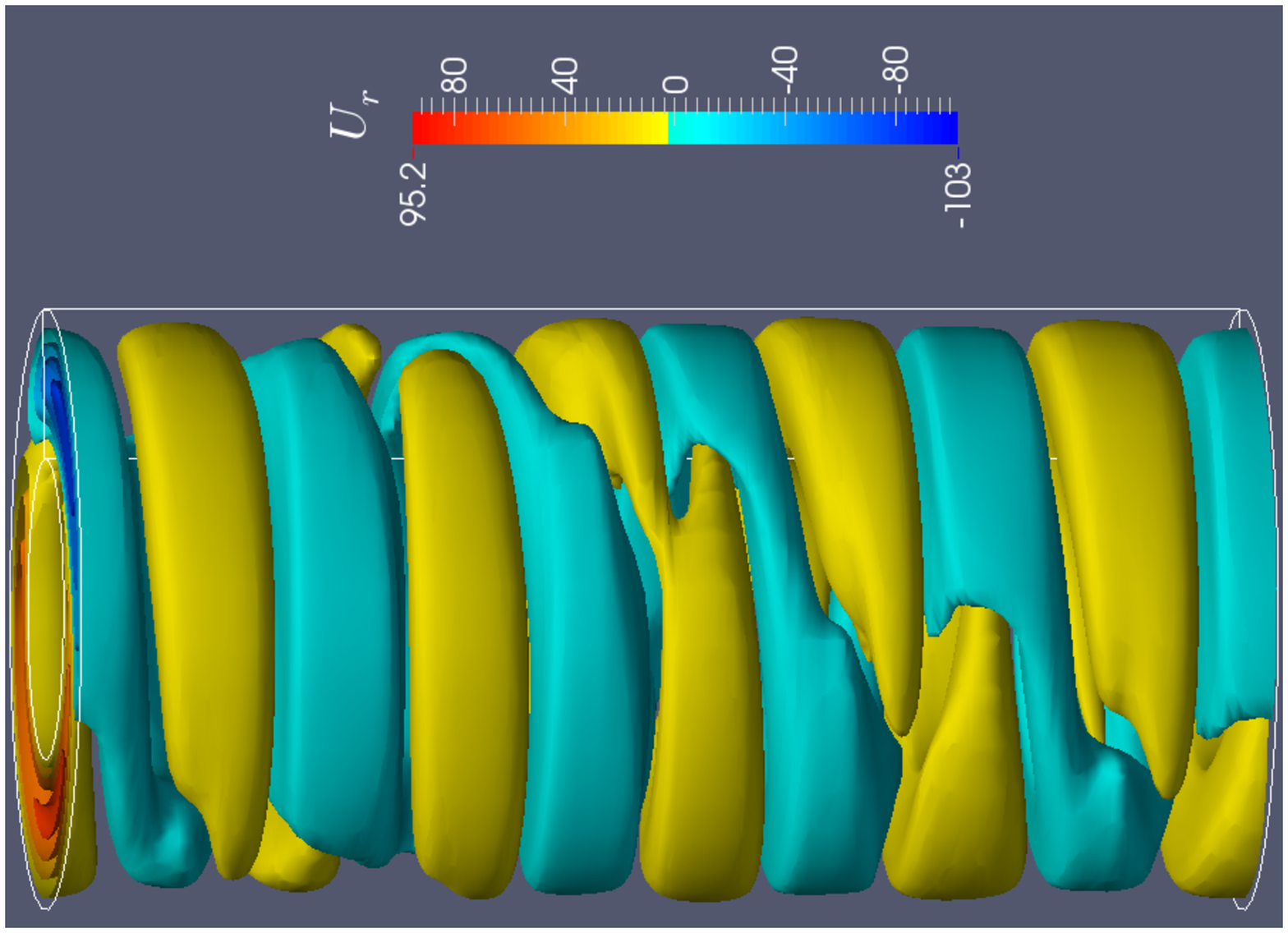}}\quad%
\subfigure[Re = 700] {\includegraphics[height=0.22\textwidth,angle=-90]{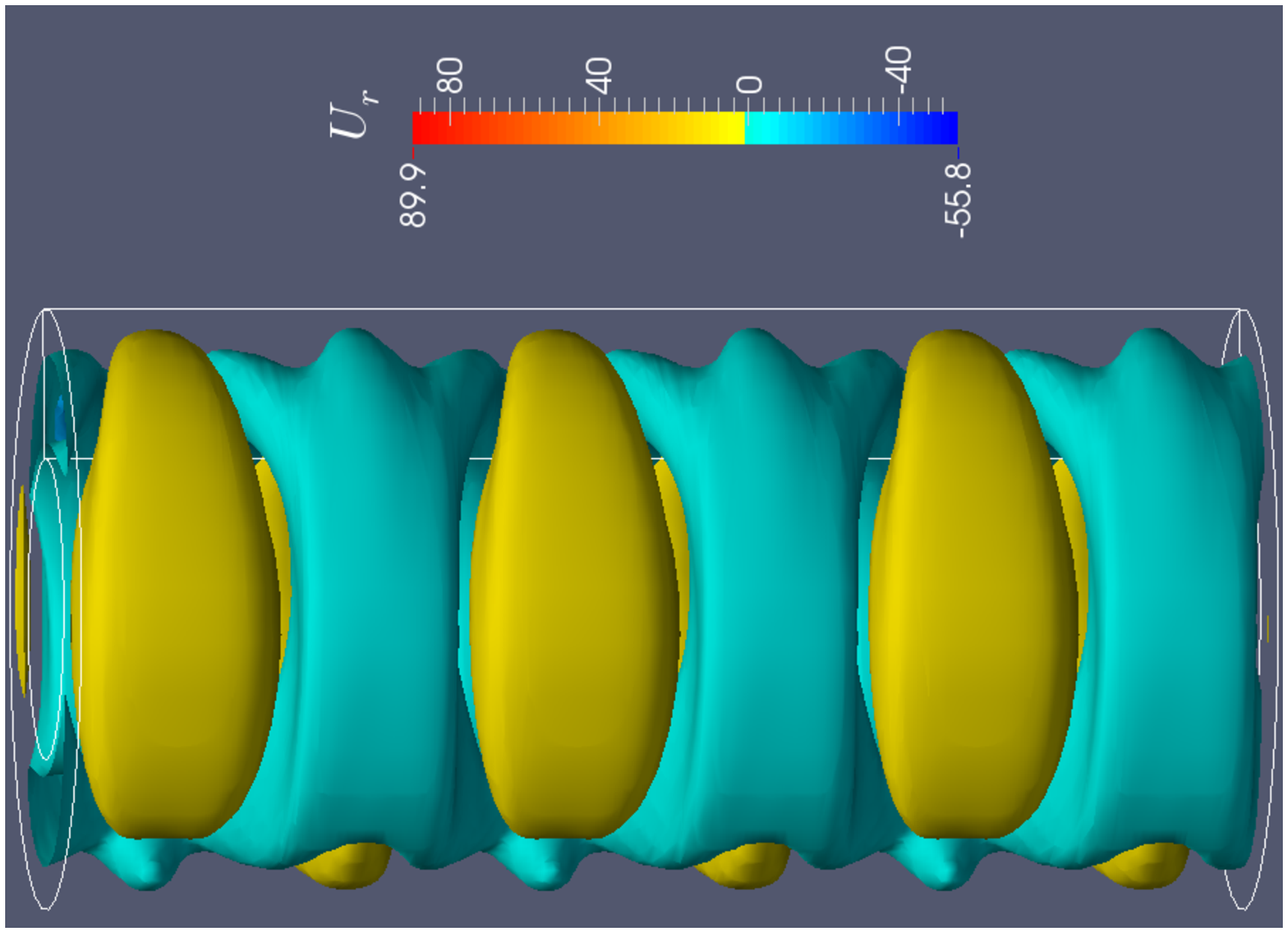}}\quad%
\subfigure[Re = 800] {\includegraphics[height=0.22\textwidth,angle=-90]{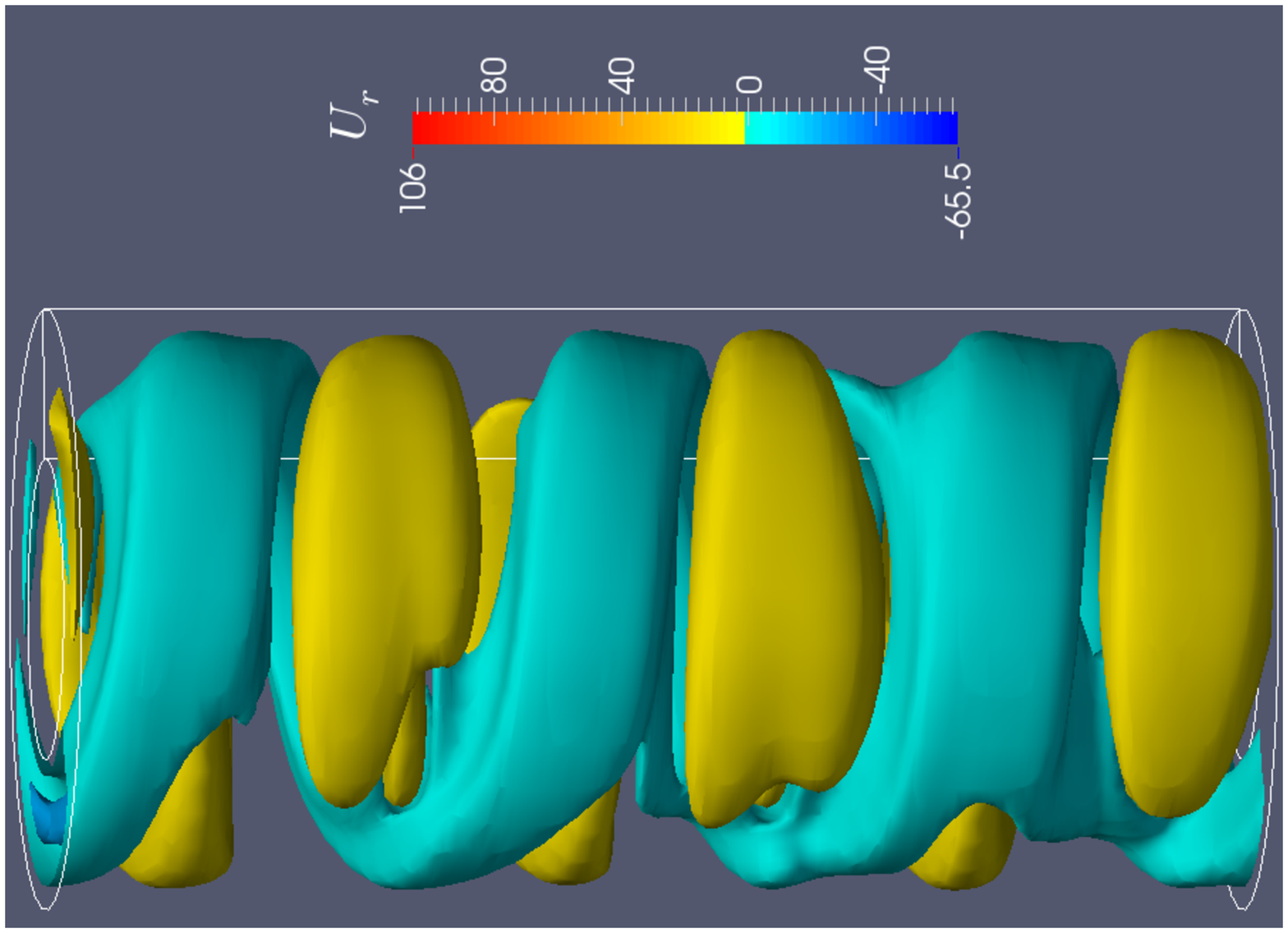}}%
\caption{Isolines for the radial component of  the velocity in units of $\nu/D$. 
 After the Taylor-Proudman theorem for faster rotation the axial wavelength  
becomes longer  and longer and the radial rms value of the velocity sinks. 
$\Ha=80$, $\mu=0.35$, $\Pm=0.1$.}
\label{iso1}
\end{figure}

The flow pattern of the  instability is shown  in  Fig.~\ref{iso1} for   
quasi-Kepler rotation of growing Reynolds numbers.  The Hartmann number is 
fixed at $\Ha=80$. The magnetic Mach number of rotation 
\be
{\rm Mm}=\frac{\Om_{\rm in}}{\Om_{\rm A}}= \frac{\sqrt{\Pm}\Rey}{\Ha}
\label{Mm}
\ee
reflects the rotation rate in units of    the \A\ frequency  ${\Om_{\rm A}= 
{B_{\rm in}}/{\sqrt{\mu_0\rho R_0^2}}}$. Almost all cosmical objects possess 
large magnetic Mach numbers. E.g., the   White Dwarfs   rotate 
with about 2 km/s while the observed magnetic field with (say) 1 MG leads to an 
\A-velocity of about 3 m/s so that $ \Mm\simeq 700$. This value even exceeds 
unity if the largest ever observed magnetic fields of 100 MG are applied.
Inserting the characteristic values for  the solar tachocline ($R_0= 1.5\cdot 
10^{10}$ cm, $\rho=0.2 $ g/cm$^3$) one finds $\rm Mm= 30/B_\phi$ with $B_\phi$ 
in kG so that with $B_\phi\lsim 1 $ kG   also the tachocline   with $\Mm\gsim 
30$  belongs to the class of rapid rotators. The upper panel of Fig.~\ref{f1} 
demonstrates that pinch models with   $\rm Mm>1$ and rigid rotation   are  
stable but they easily become  unstable  if they rotate  differentially  (see 
Fig.~\ref{f1}, bottom).  This is an important point  in the following 
discussion.

\begin{figure}[h!]
 \centering
\includegraphics[height=0.30\textwidth]{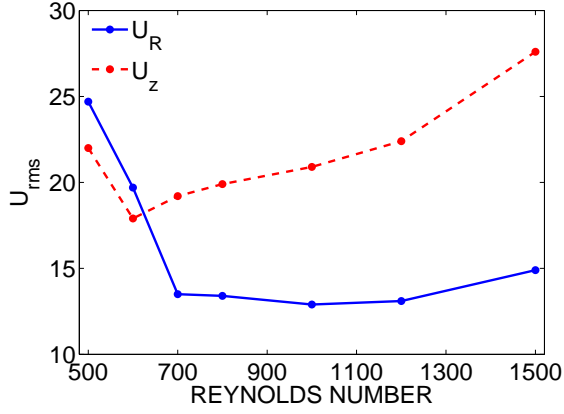}
\caption{The radial and the axial rms velocity components $u_R$ and $u_z$ for $\Mm\lsim 6$.  $\Ha=80$, $\mu=0.35$, $\Pm=0.1$. }
\label{urms}
\end{figure}

The plots of Fig.~\ref{iso1}  represent   the radial velocity which basically 
produces the radial mixing. The instability is nonaxisymmetric,  the velocity 
amplitude does hardly vary for fast rotation but the rms velocity of $u_R$ 
decreases by a factor of 1.6 between $\Rey=500$ and $\Rey=700$ reaching a 
saturation  value while the axial flow perturbation starts to rise (Fig.~\ref{urms}).  Under the influence of fast rotation a turbulence field which is 
isotropic in the laboratory system becomes  more and more anisotropic reaching a 
relation $\langle u_z^2\rangle\simeq \langle u_R^2\rangle+\langle 
u_\phi^2\rangle$ for the volume-averaged velocity. One finds from 
Figs.~\ref{iso1} and \ref{urms} that  the anisotropy -- or, with other words, 
the transition from slow rotation to fast rotation -- starts at $\Rey\approx 
600$ or $\Mm\simeq 2$, respectively.

This statement is supported by the behavior of the axial wavelength. Within the same interval also the axial wavelength -- which after the 
 Taylor-Proudman theorem should grow for  faster rotation --  seems to jump by 
the same factor. The question will be whether for Reynolds numbers of about  600 
also the diffusion coefficient jumps.

It is interesting that  the estimation $ u_{R,\rm rms} L/3$ for any  kind of 
turbulent diffusivity leads to the maximum  value  $\approx 10 \nu$ for the 
diffusion coefficient. This rather small value does {\em not} fulfill the 
constraints by Schatzman and Lebreton \& Maeder described above. 
The nonlinear simulations will show whether this preliminary result is confirmed 
or not.

\section{The diffusion equation} 
For the  Tayler unstable system the eddy diffusion of a passive scalar in radial direction is computed.  To this end, the  additional dimensionless  transport equation  
\be
\frac{\partial C}{\partial t} + \nabla \cdot \left(\vec{U} C \right) =\frac{1}{\Sc} \Delta C  \label{TEPS} 
\ee
for a passive scalar, $C$, is added to the equation  system (\ref{mhd}).  This 
passive scalar can be the temperature or  a  concentration function of chemicals 
like lithium or beryllium.  Here the microscopic  Schmidt number 
\be
{\rm Sc}=\frac{\nu}{D} 
\label{Sc}
\ee
is used in Eq.~(\ref{TEPS}), where  $D$  is the molecular  diffusivity  
of the  fluid.  The Schmidt number for gases is of order unity while it is $O(100)$ for fluids. Gough (2007) gives 
$\Sc\simeq 3$ with a (radiative) viscosity of 27
cm$^2$/s for the plasma of the solar tachocline.  In the present paper the 
molecular Schmidt number is varied from $\Sc=0.1$ to $\Sc=2$. In their 
simulations with a driven turbulence probing the Boussinesq type of the  
diffusion process Brandenburg et al. (2004) are also using $\Sc=1$ for the 
molecular Schmidt number.

When the  instability is completely developed, it will influence the transport properties of the fluid. 
This influence might be isotropic or anisotropic, thus different in radial and axial direction.  For the latter Nemri et al. (2012) and Akonur \& Lueptow (2002) find a linear dependence  between $D$  and $\Rey$ for the hydrodynamic system with resting outer cylinder.  We focus on the 
radial direction. If $D$ is considered  as the molecular  diffusivity, its modification can be modeled by an effective  diffusivity 
\be
D_{\rm eff} = D + D^{*},
\ee
where $D^*$ is only due to the magnetic-induced instability. The goal is to compute the ratio $D^{*}/D$ as a function of $\rm Re$ and $\Ha$.   When averaging Eq.~(\ref{TEPS}) 
along  the toroidal  and azimuthal directions, the mean-field  value ${\bar C}$ follows 
\be
\frac{\partial {\bar C}}{\partial t} + \nabla \cdot \left(\vec{\bar U} {\bar C }  \right) = \frac{1}{\Sc_{\rm eff}}\Delta {\bar C},
\ee 
with the effective Schmidt number $\textrm{Sc}_{\rm eff}  = {\nu}/{D_{\rm eff}}$. 

To measure the ratio $D^{*}/D$, two steps are  followed. First, a numerical 
simulation of Eqs.~(\ref{mhd}) is performed until the instability is fully 
developed and energy saturation is reached.  The saturation is achieved when the magnetic and kinetic energy of 
each mode is saturated.  Second,  the transport equation (\ref{TEPS}) is 
switched on and several simulations with different $\Sc $ numbers are performed. 
 This two steps  are repeated for several $\rm{Re}$ while all other parameters
remain  fixed.

Since the diffusion leads to a  homogenization of the passive scalar profile, the quantity $\bar C $  will exponentially decay in a characteristic
time $\tau$ which is directly related to the effective Schmidt  number.  This process will occur whether the magnetic-induced instability is present or not.  
It will be considered two decay times, i.e.  $\tau^{*}$ and $\tau$.  The first is computed from a simulation where  the  instability is present and  the second is computed
from a simulation where $\bar C $ evolves alone.  Both decay times  can inversely be related to their  diffusivity, i.e. 
\be
 \frac{D^*}{D}=\frac{\tau}{\tau^*} -1. \ee
To compute  the decay time,  the maximum  of the radial profile $\bar C $  is plotted at fixed time steps.  The characteristic time ${\tau}$ is the e-folding of the resulting profile.

\section{Results}

Numerical simulations are carried out in a Taylor-Couette container with periodic boundary conditions in the axial direction. Additionally to the boundary
conditions for the velocity and the magnetic field, Neumann boundary conditions at both $R_{\rm{in}}$ and $R_{\rm{out}}$ are  applied for the passive scalar $C$.
In order to focus only on the radial transport, the initial condition for the passive scalar $C_{0}$  is chosen to be axisymmetric and constant in the $z$-direction.
These characteristics on $C_{0}$ will allow the evaluation of the increment on the diffusivity in the radial direction.  Thus $C_{0}$ is taken  as a  radius-dependent 
Gaussian centered at the middle of the gap, i.e.
\be
 C_0 = \exp\left(-\left(\frac{r-r_{0}}{0.1}\right)^{2} \right) 
\label{Theta}
\ee
with  $ r_0 = 0.5(R_{\rm in} +  R_{\rm out})$. Since the boundary conditions are periodic in the $z$-direction and Neumann at cylinder walls, the initial condition will  evolve towards a constant profile in 
the entire cylinder.  The maximum of $\bar C $ at each time step is plotted and the characteristic decay time can be computed.
\begin{figure}[h!]
\centering
\includegraphics[width=0.38\textwidth]{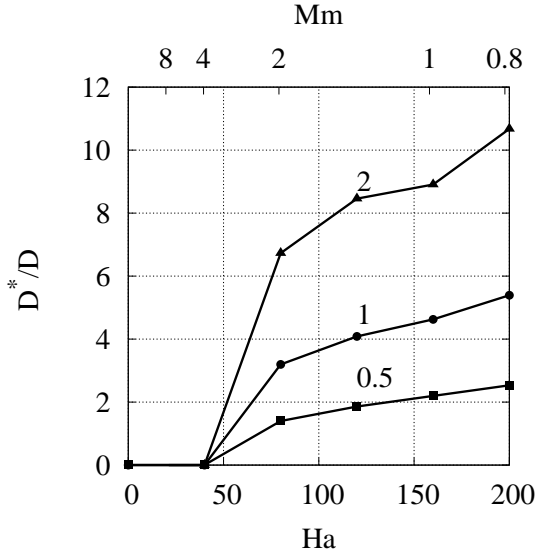}
\caption{The normalized  diffusivity $D^*/D$  vs. the  Hartmann number (bottom horizontal axis) and the magnetic Mach number (top horizontal axis). The 
curves are marked with the values of the Schmidt  number $\Sc$. The value $\Mm=1$ separates the regimes of slow and fast rotation. $\Rey=500$, $\mu=0.35$, $\Pm=0.1$.}
\label{figha}
\end{figure}

We mainly work  with a quasi-Kepler rotation profile which is unstable for the given value of $\Rey=500$ (Fig.~\ref{figha}). 
The threshold value for the onset of the instability is $\Ha\simeq 35$  from which value on the effective diffusivity grows. 
At $\Ha=158$ the magnetic Mach number $\Mm$ becomes unity defining the regimes of slow rotation ($\Mm < 1 $) and fast rotation ($\Mm > 1 $).  
In the slow rotation regime, the normalized diffusivities grow with growing $\Ha$ while they sink with decreasing $\Ha$ in the fast rotation regime
 where the rotation is fast compared to  the magnetic field. Figure \ref{figha}  also demonstrates that  $D^{*}/D$ linearly increases for increasing 
$\Sc$ so that  simply $D^{*}\propto \nu$ results  for $\Sc>0.1$.  It is the molecular  viscosity  alone which determines  the diffusion effect of
 the magnetic-induced instability -- just in the sense of Schatzman.

\begin{figure}[h!]
\vspace{0.5cm}
\centering
\includegraphics[width=0.38\textwidth]{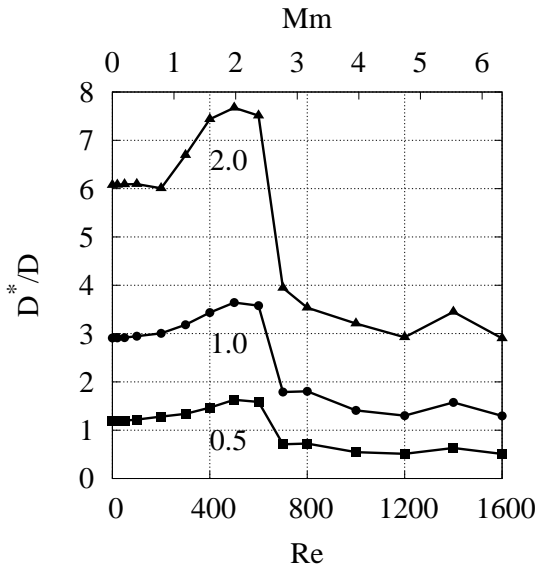}
\caption{The same as in Fig.~\ref{figha} but  in relation to the Reynolds number (bottom horizontal axis) and the magnetic Mach number 
(top horizontal axis) for $\Ha=80$. The value $\Mm=1$ separates the regimes of slow and fast rotation. Note the  reduction 
of $D^*/D$ for $\Mm> 2$. ${\mu=0.35}$, ${\Pm=0.1}$.} \label{f4}
\end{figure}

The resulting  ratio $D^{*}/D$ for quasi-Kepler rotation and for a fixed Hartmann number  is shown in Fig.~\ref{f4}.  Now the magnetic Mach number
becomes unity at $\Rey=253$. In the slow rotation regime, the effective diffusivity hardly changes. The fact that even without rotation 
the ratio $D^{*}/D$ is different from zero is because the flow is unstable even for $\Rey=0$.
In the  fast rotation regime,  the ratio  $D^{*}/D$  increases in a monotonic way until it peaks at about $\Mm\simeq 2$. Finally, for  faster rotation ($\Mm>2$) the 
effective diffusivity decays due to the rotational quenching.  In all cases, however, the effective normalized 
diffusivity grows with growing Schmidt number so that  also here  simply  $D^{*}\propto \nu$ without any influence of the microscopic diffusivity. The 
missing factor in this relation is simply given by the curve for  $\Sc=1$ in 
Fig.~\ref{f4}. 


For all Schmidt numbers, the ratio $D^{*}/D$ increases monotonically until a certain $\rm{Re}$ is reached,  
beyond it decreases.   This behavior can be understood by the fact that  the ratio 
$D^{*}/D$ must be  a direct function of the radial velocity magnitude and the wave number of the solution,  since  $D^{*}$ is produced solely by
the instability which  modifies the radial velocity  magnitude and the   wave 
number.  As shown  in Fig.~\ref{iso1} the magnitude of the radial velocity 
component hardly changes from $\rm{Re}=400$ to $\rm{Re}=800$ while   the 
wavelength  increases between $\rm{Re}=400$ and  $\rm{Re}=700$. Thus the decreasing of the ratio $D^{*}/D$  is due to the decreasing of the wave number. 
\begin{figure}[h!]
\centering
\includegraphics[height=0.47\textwidth,angle=-90]{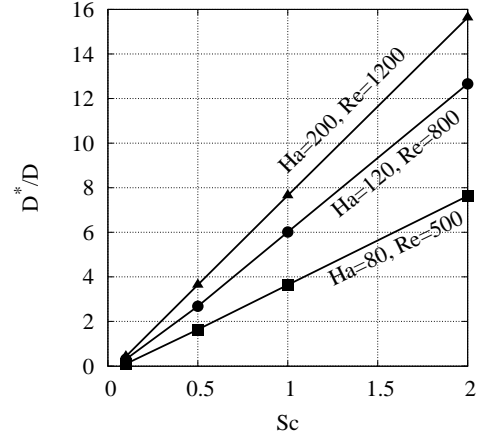}
\caption{ $D^{*}/D$ as a function of $\rm{Sc}$  for those Reynolds numbers  yielding the maximal instability-induced diffusivities. The magnetic Mach number slightly 
exceeds  2 in all  cases.  ${\mu=0.35}$,  ${\Pm=0.1}$.}
\label{ratvspr}
\end{figure}

Figure~\ref{ratvspr} shows the normalized diffusivity for $\mu=0.35$ for  fixed $\Ha$    as a function
of $\rm{Sc}$.   For all Reynolds numbers the induced  diffusivities $D^{*}$ are 
different from zero and the ratio $D^{*}/D$ scales linearly with  $\rm{Sc}$.  
The figure only shows the relation for those Reynolds numbers for which the 
maximum diffusivities for  magnetic Mach numbers about 2  are reached.
For  $\rm{Sc}\to 0$  the  diffusivity $D^{*}$ appears to vanish. 
Hence, for molecular Schmidt numbers $\Sc>0.1$ the essence of Fig. \ref{ratvspr} is the linear relation between $D^*/D$ and $\Sc$. 
In the notation  of Schatzman (1977) it means that 
\be
D^* = Re^* \ \nu
\label{schatz}
\ee
with the scaling factor $Re^*$ (which indeed forms  some kind of    a Reynolds number). Also the Figs. \ref{figha} and \ref{f4} demonstrate that  linear relations hold for all considered $\Rey$ and $\Ha$. 
The Schatzman factor  $Re^*$  after  Fig. \ref{ratvspr} {\em grows with growing} 
$\Ha$ while for all three models the magnetic Mach number is nearly the same.  
We find $Re^*\lsim 4$ for $\Ha=80$ growing to  $Re^*\lsim 8$ for $\Ha=200$; a 
saturation for larger $\Ha$ is indicated by the results presented in  Fig. 
\ref{ratvspr}. 
It is not yet clear whether for large $\Ha$ an upper limit exists for $Re^*$ due to numerical limitations.

\section{Conclusions}
The influence of the current-induced  instability on the  effective diffusivity 
in radial direction  of a rotating pinch has  been studied for different 
rotation laws. The diffusion equation is numerically solved  in a cylindric 
setup under the influence of stochastic fluctuations which are due to the 
magnetic Tayler instability. The conducting fluid between two rotating 
cylinders  becomes unstable if an axial uniform electric-current is strong 
enough. The rotation law between the two cylinders is fixed by their rotation; 
our main application is a quasi-Kepler rotation which results when the 
cylinders are rotating like planets. The magnetic Prandtl number of the 
fluid is fixed to the value of 0.1 while its Schmidt number (\ref{Sc}) 
is a free parameter of the model.

The main result is a strictly linear  relation between the resulting normalized 
eddy diffusivity $D^*/D$ and the given Schmidt number.  The Schatzman relation 
(\ref{schatz}) -- which also describes our  result that for small Schmidt number 
the eddy diffusivity is negligibly small -- has thus been confirmed in a 
self-consistent way. 

The model also provides numerical values  for the scaling factor $Re^*$ which 
increases for increasing Hartmann number of the toroidal field.   For Schmidt numbers 
$\lsim 0.1$ the diffusivity due to the magnetic instability is negligibly small,  however  
$Re^*$ is different from zero and even exceeds unity in our computations.  Already for the value $\Ha=80$ one finds 
$Re^*\lsim 4$ for quasi-Kepler rotation and this value increases for increasing Hartmann number (Fig. \ref{ratvspr}).

The second result concerns the role of the  magnetic Mach number $\Mm$ which 
represents the global rotation in relation to the magnetic field strength. For 
slow rotation the eddy diffusivity runs linear with $\Mm$ but in all cases  a 
maximum of $D^*/D$ exists  at $\Mm\simeq 2$. For faster rotation the induced diffusion 
is suppressed and finally seems to remain constant  (Fig. \ref{f4}). This fast-rotation 
phenomenon may be a consequence of the Taylor-Proudman theorem after which  the 
axial fluctuations are favored in expense of the radial ones. Also the 
correlation lengths  in axial direction grow for growing Reynolds numbers. Both 
consequences of the Taylor-Proudman theorem appear to retard  the growth of the 
radial diffusion in stars. 
 \begin{figure}[h!]
\centering
\includegraphics[height=0.5\textwidth,angle=-90]{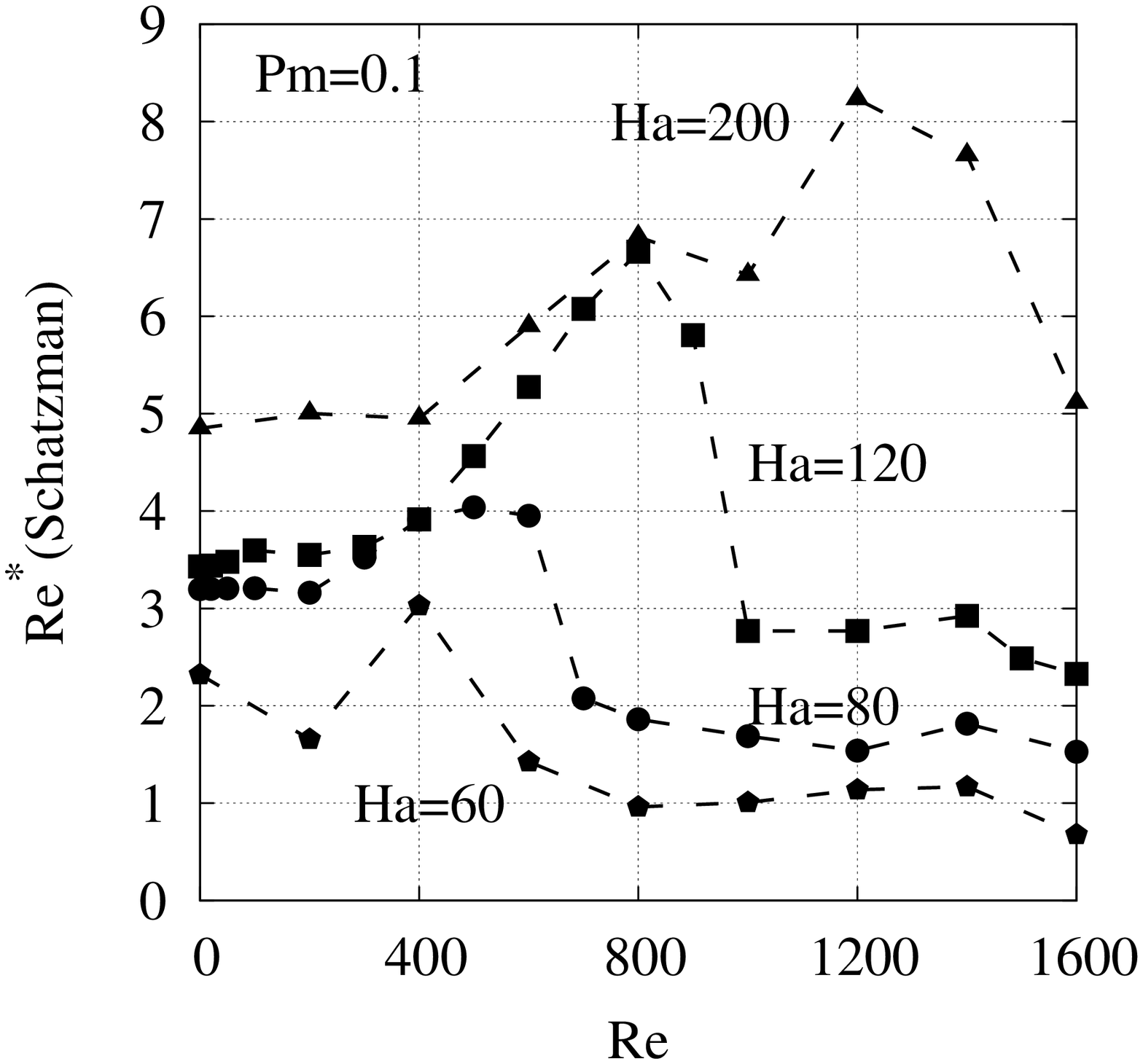}
\includegraphics[height=0.5\textwidth,angle=-90]{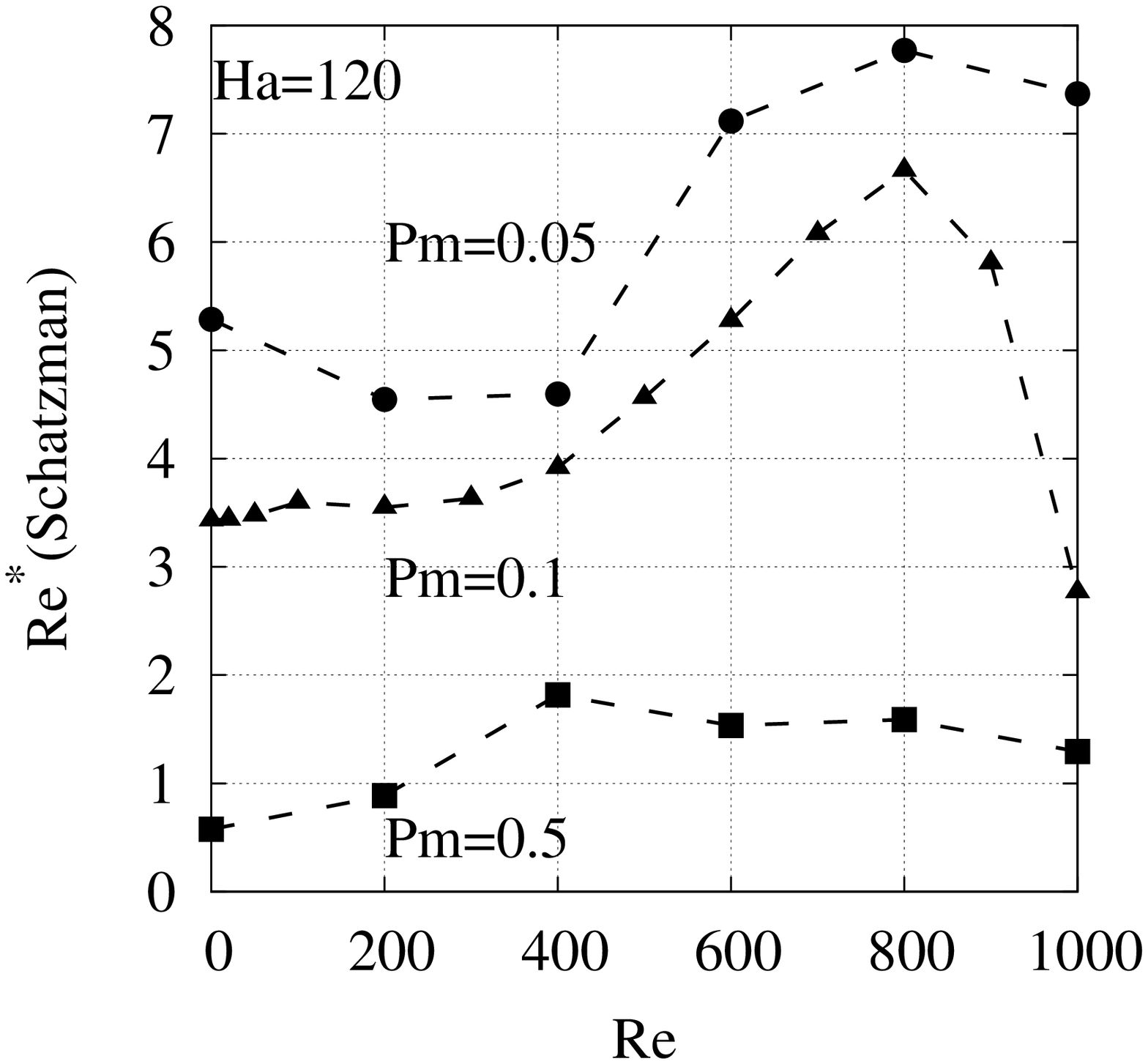}
\includegraphics[width=0.5\textwidth]{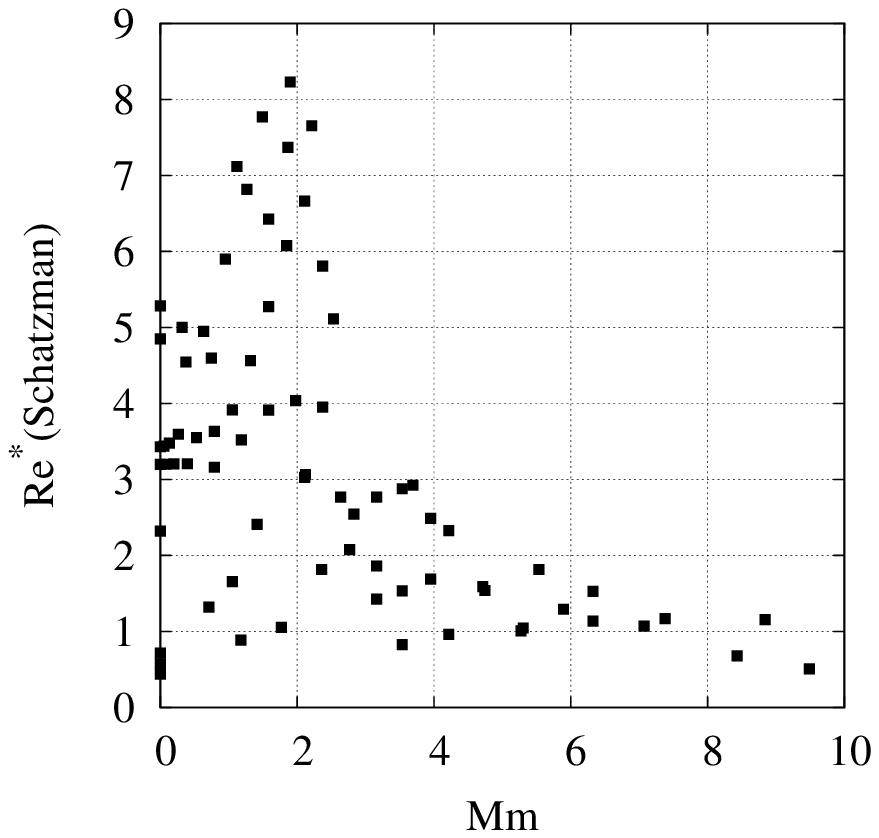}
\caption{The Schatzman number $Re^*$ vs. $\Rey$ as  functions of $\Ha$ (top panel), of $\Pm$ (middle panel) and of $\Mm$ (bottom panel).
In the bottom panel all the simulations of  this study are shown. $\mu=0.35$.}
\label{Schatzman}
\end{figure}

Figure \ref{Schatzman} summarizes the results  of this study by presenting  
numerical values of $Re^*$ for Kepler rotation laws.   In the upper panel  
$Re^*$ is given for four values of the Hartmann numbers as a function of the Reynolds 
number.  Without rotation one finds that $Re^*\propto \Ha$ is realized which for 
solar/stellar $\Ha$-values would produce even  higher $Re^*$-values  that the
ones found in our computations.  With rotation, a  maximum
of $Re^*$ exists for approximately one and the same magnetic Mach number 
($\Mm\simeq 2 $) where  the value of $Re^*$ strongly increases with the 
Hartmann number. The higher the Hartmann number the larger  $Re^*$,  but this relation is not  strictly  linear  as a mild  
saturation  may exist. For $\Mm\gg 2$ the rotational quenching  leads to
$Re^*$ even smaller than the values for $ \Rey=0$.

The numerical restrictions of our code prevent   calculations for    
higher Reynolds numbers but we are able to  vary the magnetic Prandtl number.  
This might  be necessary  as $\Pm$ in the solar tachocline is certainly 
smaller than 0.1. Figure  \ref{Schatzman} (middle panel) shows the clear 
result  that $Re^*$ is anticorrelated with the magnetic Prandtl number. The  
smaller $\Pm$ the larger $Re^*$, without rotation there is even $Re^*\propto 
1/\Pm$. For fast rotation  the results do not exclude the possibility that a 
saturation may occur  for $\Pm\ll 0.1$  so that the influence of $\Pm$ becomes 
weaker as  the close lines of marginal instability in Fig.~\ref{f1} for small $\Pm$ suggest.

The bottom panel completes the picture showing $Re^*$ as a function of $\Mm$ for all simulations 
used in this study.  It shows that  $Re^*$ has a maximum at $\Mm\simeq 2$ and  rapidly decreases 
for large $\Mm$ for which it seems to saturate around $Re^{*}=1$.

A final answer of how large the Schatzman number $Re^*$   may become by the 
presented  mechanism of  nonaxisymmetric instabilities of azimuthal fields can 
not yet be given, mainly due to  numerical limitations.  All the used dimensionless 
numbers such as the Reynolds number and the magnetic Mach number   are  
different from the real (solar) numbers by orders of magnitudes. That the model, proposed 
in this paper to estimate the magnetic induced extra diffusivity, 
directly  leads to the  original formulation of  Schatzman seems to be  a  
highly motivating  result. Using the maximal $Re^*$ values of our curves (which hold 
for $\Mm\simeq 2$) and by rescaling to the real solar/stellar Hartmann numbers 
the resulting $Re^*$-values  would become much too high. The curves in Fig. 
\ref{Schatzman}, however, demonstrate a dramatic rotational suppression of the 
diffusion process for higher $\Mm$ so that the small values $Re^*\simeq 100$ 
which are necessary to explain the observations may easily result from the 
intensive  quenching by   the global rotation belonging to magnetic Mach numbers 
exceeding (say) 30. A theoretical explanation of the slow diffusion effects 
with magnetic instabilities, therefore, requires differential  rotation and weak 
fields (of order kG) as otherwise the mixing would be too effective.
\begin{figure}[h!]
\centering
\includegraphics[height=0.37\textwidth,angle=-90]{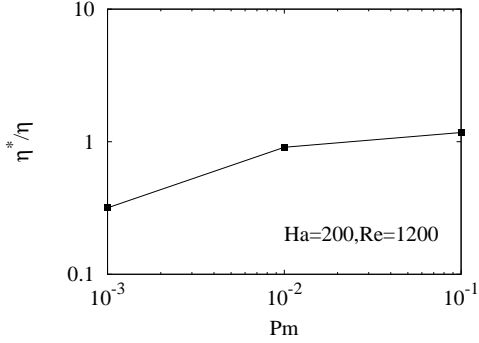}
\caption{The normalized instability-induced diffusivity $\eta^{*}/\eta$ for the differentially rotating pinch  as a function of $\Pm$  for the Reynolds  and Hartman number yielding the maximal  $Re^*$. The magnetic Mach number is about   2 (see Fig. \ref{Schatzman}). ${\mu=0.35}$.}
\label{eta}
\end{figure}

{ The presented instability model bases on the simultaneous existence of differential rotation and 
toroidal magnetic field. It will thus finish after the decay of one of the two ingredients. 
The question is which of them decays faster by the instability-induced diffusion. Provided  the 
characteristic scales of the differential rotation and the magnetic field are of the same order
(as it is the case for the magnetized Taylor-Couette flows) then the ratio of the decay times
of the magnetic field and the differential rotation is 
\be
\frac{\tau_{\rm mag}}{\tau_{\rm rot}} = \frac{\nu+\nu^*}{\eta+\eta^*}. 
\label{ratio}
\ee
As the angular momentum transport is also due to the Maxwell stress of the fluctuations 
the turbulent viscosity always considerably exceeds the molecular viscosity which -- for small $\Pm$ -- 
is not  the case for the magnetic resistivities. We always find for the rotating pinch the $\eta^*$ to be of order of $\eta$ (Fig. \ref{eta}). The instability-induced $\eta^*$ result from the  defining relation  $\eta^*=\langle u_\phi b_R-u_Rb_\phi\rangle/2 A$ of the instability-originated axial component of the electromotive force $\langle \vec{u} \times \vec{b}\rangle$ (\R\ et al. 2016).

At least for small magnetic Prandtl number, therefore,  the instability does not basically   accelerate the decay of the fossil magnetic field. Hence, $\tau_{\rm mag}/\tau_{\rm rot} \propto {(\nu^*}/{\nu})\,\Pm$.  This expression decreases with decreasing magnetic Prandtl number as also the  normalized turbulent viscosity $\nu^*/\nu$ sinks for decreasing $\Pm$ for fixed  Reynolds number (R\"udiger et al. 2015b). For  small magnetic Prandtl number, therefore, the differential rotation   {\em never} decays faster than the magnetic field which by itself   decays at the  long microscopic diffusive  timescale.  Only for large magnetic Prandtl number the magnetic angular momentum transport might stop the instability prior to the decay of the fossil field.}

\begin{acknowledgements}
 This work was supported by  the Deutsche
Forschungsgemeinschaft within the SPP Planetary Magnetism.
\end{acknowledgements}



\bibliographystyle{plain}
{}

\end{document}